\documentclass{article}
\usepackage{amsmath,graphicx}
\usepackage{xcolor}

\usepackage[preprint]{spconf}
\title{SIG-VC: A Speaker Information Guided Zero-shot Voice Conversion System for both Human Beings and Machines}
\name{Haozhe Zhang$^{1}$, Zexin Cai$^{1}$, Xiaoyi Qin$^{1,2}$, Ming Li\sthanks{Corresponding author: Ming Li}$^{1,2}$ }
\address{$^{1}$Data Science Research Center, Duke Kunshan University, Kunshan, China\\
$^{2}$School of Computer Science, Wuhan University, Wuhan, China\\}
\email{ming.li369@duke.edu}
\copyrightnotice{\copyright\ 2022 IEEE Accepted by ICASSP 2022}
\begin{document}
\ninept
\maketitle
\begin{abstract}
Nowadays, as more and more systems achieve good performance in traditional voice conversion (VC) tasks, people's attention gradually turns to VC tasks under extreme conditions. In this paper,  we propose a novel method for zero-shot voice conversion. We aim to obtain intermediate representations for speaker-content disentanglement of speech to better remove speaker information and get pure content information. Accordingly, our proposed framework contains a module that removes the speaker information from the acoustic feature of the source speaker. Moreover, speaker information control is added to our system to maintain the voice cloning performance. The proposed system is evaluated by subjective and objective metrics. Results show that our proposed system significantly reduces the trade-off problem in zero-shot voice conversion, while it also manages to have high spoofing power to the speaker verification system.
\end{abstract}

\begin{figure*}[htbp]
  \centering
  \includegraphics[scale=0.4]{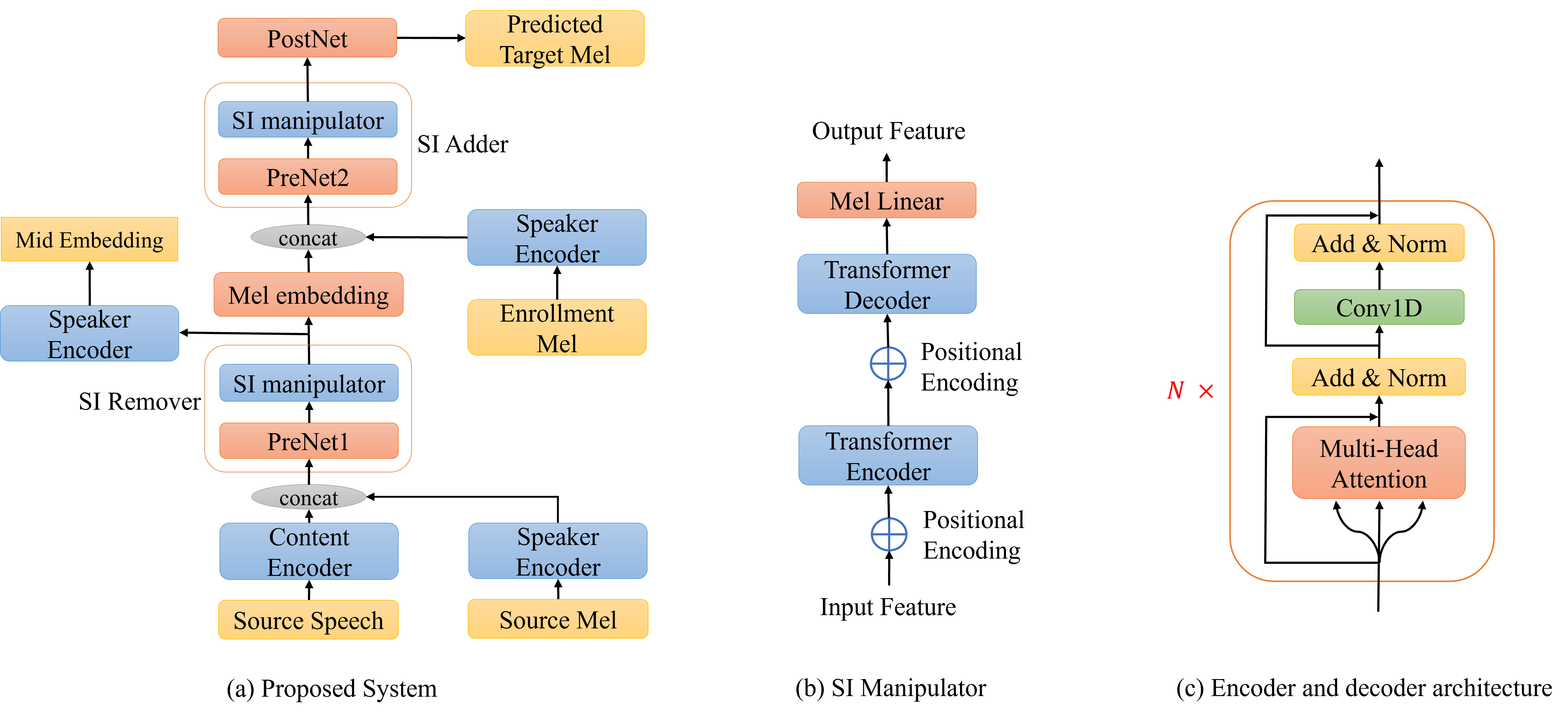}
  \caption{The structure of our proposed system, SI stands for speaker information}
  \label{fig:architecture}
\end{figure*}

\section{Introduction}
\label{sec:intro}

Voice Conversion (VC) is a task that aims at converting the timbres from source speakers to target speakers while keeping the content of speech unchanged. For traditional VC tasks, many systems manage to synthesize speech with good quality, in which the source and target speakers are seen in the training data. On the other hand, it remains a challenge for voice conversion under the condition that the data of source or target speakers is scarce and unseen in the training stage. Zero-shot voice conversion is one of the challenges, and most VC systems are not robust enough for it.


Previously, there are research works proposed for addressing the unseen voice conversion problem. In the perspective of frameworks, generative adversarial networks (GAN) and auto-encoder are the most frequently used ones. StarGAN is a GAN-based voice conversion framework proposed in \cite{stargan-origin}, then \cite{stargan, gazev} extended this framework to achieve one-shot VC. \cite{convoice, autovc, adain, again, iclr, ppp, global, sv, vqvc+} are all auto-encoder-based (AE-based) systems, with different variants. In particular, \cite{iclr} proposes an AE-based system to extract the speaker embedding and content embedding using mutual information. Alternatively, \cite{vqvc+} adopts the U-net architecture in AE-based VC, along with vector quantization \cite{vq} to disentangle the speaker information and the content information. In addition, \cite{adain, again} disentangle the speaker and content information by instance normalization, which is also used in \cite{vqvc+, gazev}. \cite{fvae} try to separate speaker and content information using variational auto-encoder (VAE). To better achieve the disentanglement capability, several works use pre-trained models as the speaker or content encoder. For instance, with the support from the automatic speech recognition system, \cite{ppp, global} use the phonetic posteriorgrams as the content representation in VC. On the contrary, \cite{autovc, sv} uses speaker verification (SV) systems as their speaker encoders to obtain discriminative representations.

Many aforementioned systems disentangle speaker and content information by finding plausible latent representations. Nevertheless, most of them apply the reconstruction loss to minimize the difference between the predicted acoustic feature and the target one during training without direct supervision on latent representations. In this case, it is difficult for latent representations to be robust enough for disentangled content or speaker information. We believe this might be the reason why the trade-off question mentioned in \cite{again} exists, which refers to the issue that the aforementioned systems are not very robust in synthesizing speech with both good similarity and naturalness under the zero-shot scenario. 

In this paper, we propose SIG-VC, a \textbf{S}peaker \textbf{I}nformation \textbf{G}uided \textbf{V}oice \textbf{C}onversion system. In our proposed system, we introduce a novel approach to supervise the intermediate representation. This supervision aims to remove the speaker information from the linguistic information. Specifically, we employ a pre-trained speaker verification system for speaker representation extraction. Regarding the linguistic content, a pre-trained acoustic model is applied to extract the linguistic feature. Experiments in \cite{yyg} show that there is speaker information persisting in the linguistic information extracted by the acoustic recognition model.  We manage to get purified content information by eliminating the residual speaker information in the speech. More detailedly, by sharing parameters for some modules in the system, we force the intermediate representation to be in the vector space of our selected acoustic feature. Then we erase the speaker information from the intermediate representation with the supervision of the speaker encoder. Finally, by conditioning on the speaker information of the target speaker, we manage to generate the target speech. Our proposed system also applies the feedback loss control on the output, which is proposed by \cite{feedback} and applied on VC in \cite{feedback-vc}. By adding this loss, the generated results of our proposed system demonstrate high spoofing capability to speaker verification systems. 

To our knowledge, we are the first to propose an approach to supervise the intermediate representation in order to remove speaker information and get pure content representation. By applying this intermediate representation supervision mechanism, our system outperforms the previous state-of-the-art system, AGAIN-VC \cite{again} in zero-shot voice conversion.




This paper is organized as follows: Section \ref{sec:method} introduces our proposed framework for zero-shot voice conversion. Section \ref{implementation} exhibits the implementation details of our proposed system. The experimental details, along with the results, are presented in Section \ref{sec:exp}. Finally, we conclude our work in Section \ref{sec:con}.  

\section{PROPOSED METHODS}
\label{sec:method}
\subsection{System overview}
\label{ssec:overview}
Figure \ref{fig:architecture} shows the structure of our proposed system. A pre-trained content encoder is applied to extract linguistic information, which contains residual speaker information. Also, a pre-trained speaker encoder is used to extract speaker embedding from the same utterance. The output of the two encoders is concatenated and fed into the PreNet1 module. The PreNet1 is designed for projecting the input into latent representations such that the following speaker information manipulator can remove the speaker information. Similarly, PreNet2 is designed for projecting the input into another latent space so that the same speaker information manipulator can add the target speaker's information to the intermediate content representation. Here, PreNet1 and SI manipulator together is called SI remover, while PreNet2 and SI manipulator together is called SI adder. The Mel embedding layer is used to project the Mel-spectrogram feature  without speaker information back into linguistic information. Finally, a PostNet is designed to finetune the predicted Mel-spectrogram.

\subsection{Content encoder}
\label{ssec:content encoder}
The content encoder in our proposed system is part of a speech recognition model trained by Kaldi \cite{Povey_ASRU2011}. The speech recognition model is constructed by time-delayed neural networks (TDNN), where the linear layer before the output layer is designed to be a low-dimensional layer, which is also known as the bottleneck layer \cite{grezl2007probabilistic}. The output of this layer is a good representation of linguistic information. Thus, we take the part before the output layer as our content encoder.

\subsection{Speaker encoder}
\label{ssec:speaker encoder}
In general, we want our speaker encoder to extract speaker information that captures the voice characteristics. Speaker representations from different speakers are expected to be distinguishable in the corresponding vector space. To that end, speaker verification (SV) systems are suitable choices. SV systems can be used to extract speaker embeddings to represent voices from different speakers. Since the encoder is trained and inferred in a text-independent manner, most of the information left in the embedding is about the speaker's timbre. Therefore, we believe the speaker embedding extracted by text-independent SV system can well represent timbres. In this case, we incorporate the ECAPA-TDNN \cite{ecapatdnn} as our speaker encoder in our proposed model.

\subsection{PreNet and speaker information manipulator}
\label{ssec:prenet and SI}
The PreNet is a stack of two linear layers, with each followed by a ReLU activation layer and a dropout layer. The structure of the speaker information manipulator is shown in part (b) and (c) of Figure \ref{fig:architecture}. This structure is a modified version of the FastSpeech \cite{fastspeech} model on VC task. The input is positionally encoded and fed into the transformer encoder or decoder with a multi-head attention mechanism. The Mel Linear layer is a linear layer that transforms the output of the decoder into the same Mel space as the input Mel-spectrogram. PreNet1, along with the speaker information manipulator, achieves the function of removing speaker information. Alternatively, PreNet2 along with the speaker information manipulator, achieves the function of adding speaker information. Since the parameters are shared for the two speaker information manipulators, PreNet1 essentially indicates the remove of speaker information while PreNet2 indicates the addition of speaker information.

\subsection{Supervision on intermediate representation}
\label{ssec:mid loss}
According to the experiment results in \cite{yyg}, although the pre-trained content encoder can provide precise linguistic information, it still contains speaker information. Thus we apply the speaker information remover as shown in Figure \ref{fig:architecture} to remove the speaker information, and ideally, we can obtain purified content information. 

Since it is hard to directly evaluate whether the speaker information is removed or not, we propose a method to force the intermediate representation to be in the same vector space as the input acoustic feature, which is plausible for extracting speaker embedding. Accordingly, we can eliminate the speaker information by minimizing the values of all dimensions of the intermediate representation's speaker embedding.

In our model, the intermediate representation refers to the output of speaker information remover. Since the two speaker information manipulators in Figure \ref{fig:architecture} share the same set of parameters, and the output of the speaker information manipulator in SI adder is supervised to be in the Mel-spectrogram space, the output of the SI manipulator in SI remover is therefore forced to be in Mel-spectrogram space as well. As our speaker encoder takes Mel-spectrograms as input and predicts corresponding speaker embeddings, we expect the speaker encoder's prediction on intermediate representation to provide speaker information. We consider the value on each dimension of the speaker embedding as an indicator for a certain characteristic of the corresponding speaker's timbre. Then minimizing the values on all dimensions of speaker embedding is one plausible way to remove speaker information. Denoting $\Vec{e}$ as the speaker embedding extracted from intermediate representation, we define a training loss called intermediate speaker loss as follows:
\begin{equation} L_{mid\_spk}=\Vert \Vec{e}-\boldsymbol{0} \Vert_1 \label{mid_spk}
\end{equation}

\subsection{Supervision on prediceted Mel-spectrogram}
During the training process, our proposed model learns to remove the speaker information from the given Mel-spectrogram and then add the same speaker information to reconstruct the original Mel-spectrogram. Let $\hat X$ denote the output of SI adder, and $X$ denote the given Mel-spectrogram. Reconstruction loss is defined as follows to make the predicted Mel-spectrogram close to the target Mel-spectrogram.
\begin{equation}
L_{recon}=\Vert X-\hat{X} \Vert_1 \label{recon}
\end{equation}
We design the PostNet to finetune the predicted Mel-spectrogram. The reconstruction loss is also applied to the output of PostNet. Denoted the output of PostNet as $\hat{X}_{postnet}$, the Postnet reconstruction loss is
\begin{equation}
L_{recon\_postnet}=\Vert X-\hat X_{postnet} \Vert_1 \label{recon_postnet}
\end{equation}

To better control the distribution of the predicted Mel-spectrogram, we also apply standard deviation loss. We denote the target Mel-spectrogram as $X$ with shape $T \times d$, where d represents the dimensionality of the Mel-spectrogram and T represents the number of frames. Then let $X_{std}$ denote a $d \times 1$ standard deviation vector where \begin{equation}
X_{std}^{i}=\sqrt{\frac{\sum_{t=0}^{T}(X_{i,t}-\bar{X_i})^2}{T}},0<i \le d \label{std}
\end{equation}
Let $\hat X_{std}$ denote the standard deviation vector for the output of PostNet. Then the standard deviation loss is defined as \begin{equation}
L_{std}=\Vert X_{std}-\hat X_{std} \Vert_1 \label{std_loss}
\end{equation}

We also apply the loss proposed by \cite{feedback} to the output of PostNet as the speaker reconstruction loss. Denoted the speaker embedding extracted from target Mel-spectrogram as $\boldsymbol{s}$ and the one extracted from the predicted Mel-spectrogram as $\boldsymbol{\hat{s}}$, then the speaker reconstruction loss is 
\begin{equation}L_{spk}=1-cos<\boldsymbol{s}, \boldsymbol{\hat{s}}> \label{spk_loss}
\end{equation} where $cos<\boldsymbol{s},\boldsymbol{\hat{s}}>$ means the cosine similarity between the vector $\boldsymbol{s}$ and $\boldsymbol{\hat{s}}$.

\begin{figure}[htbp]
  \centering
  \includegraphics[scale=0.33]{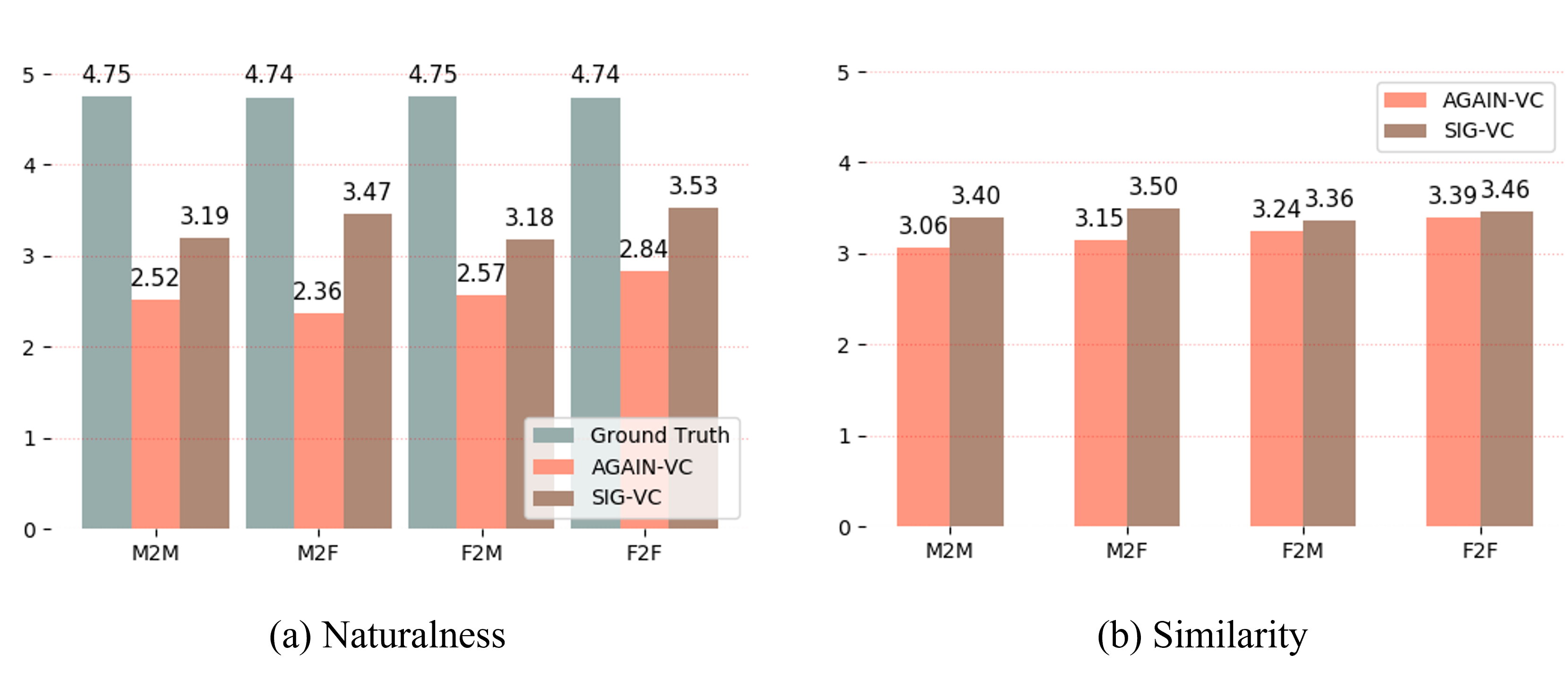}
  \caption{The Mean Opinion Score for AGAIN-VC and our proposed system. M for male and F for female. M2F means the source speaker is male and the target speaker is female}
  \label{fig:mos}
\end{figure}

\subsection{Training loss}
In the training stage, we use the following loss,
\begin{equation}
L = L_{mid\_spk}+L_{recon}+L_{recon\_postnet}+L_{std}+\lambda \cdot L_{spk}
\end{equation}
where $\lambda$ is the coefficient for speaker loss. 

\section{IMPLEMENTATION DETAILS}
\label{implementation}

\subsection{Data processing}
We use 16000Hz as the sampling rate to re-sample from the original dataset. Also, we trim the silence at the beginning and the end of each audio clip.

\subsection{Mel-spectrogram extraction}
The Mel-spectrogram is extracted from the audios with 80 mel bins, 256 hop length and 1024 window size for Short-time Fourier Transform (STFT), corresponding to the configuration in MelGAN \cite{melgan}.

\subsection{Training details}
Our proposed model is trained with ADAM optimizer, with learning rate of 0.001, $\beta_1=0.9$, $\beta_2=0.98$ \cite{adam}. During training, the batch size is set to 16. $\lambda$ is set to be 3.

\subsection{Vocoder}
We use MelGAN \cite{melgan} to reconstruct the time-domain waveform from Mel-spectrogram. We reproduce the MelGAN network using the official open-source code on Github\footnote{https://github.com/descriptinc/melgan-neurips}.

\section{EXPERIMENTS}
\label{sec:exp}
We conduct voice conversion experiments for unseen source and target speakers on our proposed system. Comparatively, we reproduce the state-of-the-art system, AGAIN-VC \cite{again}. The VCTK dataset \cite{veaux2016superseded} is used for training and test. For the test set, we first select the 4 speakers who appear on the demo page of AGAIN-VC, among which there are 2 males and 2 females. Then we randomly select another 3 male and 3 female speakers for the zero-shot evaluation. Utterances from the remaining 99 speakers are used as the training set. We use the same training set to train our proposed system and the AGAIN-VC. The quality of the generated samples of our trained AGAIN-VC system is close to the ones shown on the demo page.

\begin{figure}[htb]
  \centering
  \includegraphics[scale=0.4]{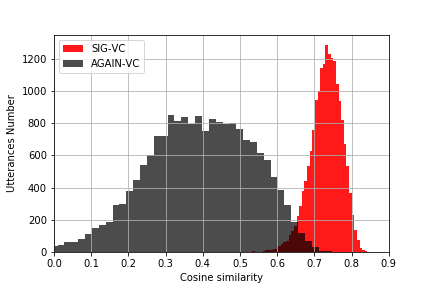}
  \caption{The distribution of the cosine similarity between the predicted speaker embedding and target speaker's average embedding extracted by ECAPA-TDNN SV system}
  \label{fig:cos1}
\end{figure}

\begin{figure}[htb]
  \centering
  \includegraphics[scale=0.4]{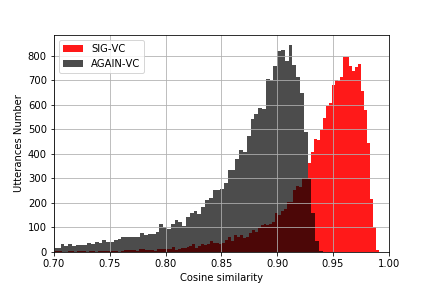}
  \caption{The distribution of the cosine similarity between the predicted speaker embedding and target speaker's average embedding extracted by ResNet34-based SV system}
  \label{fig:cos2}
\end{figure}

\subsection{Subjective evaluation}
To evaluate the naturalness and similarity of synthesized speech, we conduct the Mean Opinion Score (MOS) evaluation and ask 17 participants to rate the utterances, with each of the participants speaks fluent English. There are a total of 180 synthesized utterances for each system. Figure \ref{fig:mos} (a) shows the naturalness score for ground truth utterances, utterances generated by AGAIN-VC, and those generated by our proposed system. We present the score based on the gender of the source and target speakers. Comparing to the AGAIN-VC system, our proposed system achieves better naturalness scores in all source-target gender pairs. The subjective similarity score is shown in Figure \ref{fig:mos} (b), which shows that the synthesized results from our proposed model also outperform those from AGAIN-VC on the similarity to target speakers. Our demo is also available online for readers\footnote{https://haydencaffrey.github.io/sigvc/index.html}. 

\subsection{Speaker embedding evaluation}
We conduct objective evaluations on speaker embedding as an additional metric to assess whether two utterances belong to the same speaker. We calculate the cosine similarity between the predicted speaker embedding and the average speaker embedding of the target speaker for both our proposed system and the AGAIN-VC. Each system has the same 18000 utterances. Figure \ref{fig:cos1} shows the distribution of cosine similarity score when the speaker verification model ECAPA-TDNN is used to extract speaker embeddings. Comparing with AGAIN-VC, there is a significant improvement in the speaker similarity from the verification system's perspective. Since we incorporate ECAPA-TDNN as the speaker encoder in our proposed system, the cosine similarity score is expected to improve when evaluated by the same model. Therefore, we also evaluate the similarity performance by another ResNet34-based SV system \cite{cai2018exploring}. In this case, our proposed system also outperforms AGAIN-VC as shown in Figure \ref{fig:cos2}. 

\begin{figure}[htb]
  \centering
  \includegraphics[scale=0.42]{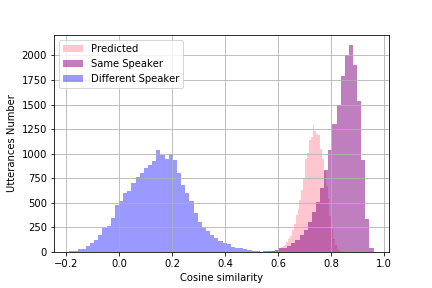}
  \caption{The distribution of cosine similarity of speaker embeddings in three conditions. }
  \label{fig:cheat_sv}
\end{figure}

We also compute the cosine similarity for three pairs of speaker embeddings in Figure \ref{fig:cheat_sv}, the similarity scores between predicted embeddings and average target embedding, the scores between the embeddings of different utterances of the same speaker and the average embedding of that speaker, and the scores between the embeddings of utterances of different speakers and the average embedding of another different speaker. The result is shown in Figure \ref{fig:cheat_sv}, which indicates that cosine similarity score of about 0.6 is a good threshold to distinguish whether the two utterances belong to the same speaker. And the distribution of the cosine similarity of our predicted utterances indicates that the synthesized utterances can be used to attack SV systems.

\section{Conclusion}
\label{sec:con}
We propose a novel approach to supervise the intermediate representation to remove the speaker information from the content information for zero-shot voice conversion. This approach greatly reduces the trade-off between similarity and naturalness of synthesized speech. Additionally, we propose the speaker information guidance mechanism that improves the speaker similarity of synthesized speech from both human's and SV systems' perspectives. The code of our proposed system is available online\footnote{https://github.com/HaydenCaffrey/SIG-VC}. 

\section{Acknowledgment}
This research is funded in part by the National Natural Science Foundation of China (62171207), the Fundamental Research Funds for the Central Universities (2042021kf0039), Science and Technology Program of Guangzhou City (202007030011). Many thanks for the computational resource provided by the Advanced Computing East China Sub-Center.





\bibliographystyle{IEEEbib}
\bibliography{mybib}
\end{document}